\documentclass[twocolumn,appendixfloats]{emulateapj}
\usepackage{float}
\usepackage{epsfig}
\usepackage{graphicx}
\usepackage{graphics}
\usepackage[latin1]{inputenc}
\usepackage{latexsym}
\newcommand{\nd}{\multicolumn{1}{c}{$\dots$}}
\newcommand{\ncep}{1417}
\newcommand{\ncpfu}{866} 
\newcommand{\ncpfo}{551} 
\newcommand{\ncper}{66}

\slugcomment{The Astronomical Journal, 149:117 (2015) and erratum}
\shorttitle{NIR Leavitt Law in the LMC}
\shortauthors{Macri et al.}

\begin{document}

\title{Large Magellanic Cloud Near-Infrared Synoptic Survey. I.\\Cepheid variables and the calibration of the Leavitt Law}

\author{Lucas M.~Macri\altaffilmark{1,*}, Chow-Choong Ngeow\altaffilmark{2}, Shashi M.~Kanbur\altaffilmark{3}, Salma Mahzooni\altaffilmark{1} \& Michael T.~Smitka\altaffilmark{1}}

\altaffiltext{1}{Mitchell Institute for Fundamental Physics \& Astronomy, Department of Physics \& Astronomy, Texas A\&M University, College Station, TX 77843, USA.}

\altaffiltext{2}{Graduate Institute of Astronomy, National Central University, Jhongli 32001, Taiwan.}

\altaffiltext{3}{Department of Physics, The State University of New York at Oswego, Oswego, NY 13126, USA.}

\altaffiltext{*}{Corresponding author; {\tt lmacri@tamu.edu}}

\begin{abstract}

We present observational details and first results of a near-infrared ($JHK_s$) synoptic survey of the central region of the Large Magellanic Cloud (LMC) using the CPAPIR camera at the CTIO 1.5-m telescope. We covered 18~sq.~deg.~to a depth of $K_s\!\sim\!16.5$~mag and obtained an average of 16 epochs in each band at any given location. Our catalog contains more than $3.5\times10^6$~sources, including $\ncep$~Cepheid variables previously studied at optical wavelengths by the OGLE survey. Our sample of fundamental-mode pulsators represents a nine-fold increase in the number of these variables with time-resolved, multi-band near-infrared photometry. We combine our large Cepheid sample and a recent precise determination of the distance to the LMC to derive a robust absolute calibration of the near-infrared Leavitt Law for fundamental-mode and first-overtone Cepheids with $10\times$ better constraints on the slopes relative to previous work. We also obtain calibrations for the tip of the red giant branch and the red clump based on our ensemble photometry which are in good agreement with previous determinations.
 
\end{abstract}


\section{Introduction}

The Cepheid Period-Luminosity relation \citep[hereafter ``The Leavitt Law'',][]{leavitt12} is one of the cornerstones of the extragalactic distance scale. It has been widely used over the past century, from Hubble's proof of the extragalactic nature of ``spiral nebulae'' \citep{hubble25} to the most accurate and precise local determination of the Hubble constant ($H_0$) to date \citep{riess11}. Increasingly more precise and accurate determinations of $H_0$ provide needed additional constraints on the equation of state of dark energy and other important cosmological parameters \citep{weinberg13}. In order to achieve these goals, further improvements in the characterization of the Leavitt Law are required. These include a more robust zeropoint calibration, better constraints on variations of zeropoint and slope as a function of metallicity, and stronger limits on nonlinearity.

The first generation of microlensing surveys directed toward the LMC resulted in the discovery of thousands of Cepheid variables \citep[][from the MACHO \& OGLE surveys respectively]{alcock99,udalski99}. At optical wavelengths, many studies have been carried out using the Cepheid photometry obtained by the OGLE survey, which provides excellent phase coverage in the standard $BVI$ bands \citep[e.g.,][]{ngeow05,bono10}. At near-infrared wavelengths, the observational material available to date is more limited. Both 2MASS \citep{skrutskie06} and the IRSF Magellanic Clouds Point Source Catalog \citep{kato07} provide complete coverage of the galaxy in $JHK_s$ but are limited to a single epoch, requiring corrections to mean light \citep{nikolaev04,soszynski05}. Synoptic observations are either limited to a single band \citep[$K_s$ for the VMC,][]{cioni11} or to a relatively small number of variables compared to the optical samples \citep[$N\!=\!92$,][hereafter P04]{persson04}.

Recently, \citet{pietrzynski13} obtained a very accurate and precise determination of the distance to the Large Magellanic Cloud through the discovery and analysis of detached eclipsing binary systems, $D\!=\!49.97\pm2\%$~kpc (equivalent to a distance modulus of $\mu_0\!=\!18.493\pm0.048$~mag). Such a robust distance estimate makes the LMC a very important component in the ``first rung'' of the extragalactic distance scale by enabling absolute calibrations of many distance indicators, such as Cepheids.

Motivated by the above, we carried out a synoptic multi-wavelength near-infrared survey of the central region of the LMC that has yielded well-sampled light curves for $\ncep$~Cepheids and an additional $\sim 3.5\times10^6$ sources. This paper, the first in a series, presents details of the observations, data reduction and photometry (\S2) and the resulting Cepheid light curves and Leavitt Law (\S3). Future work will include a Fourier analysis of Cepheid light curve structure \citep{bhardwaj15}, a study of nonlinearity in the Leavitt Law and P-L relations of long-period variables, among other topics.

\section{Observations, data reduction,\\photometry \& calibration}

\subsection{Observations and data reduction}

\begin{figure*}
\begin{center}
\includegraphics[width=\textwidth]{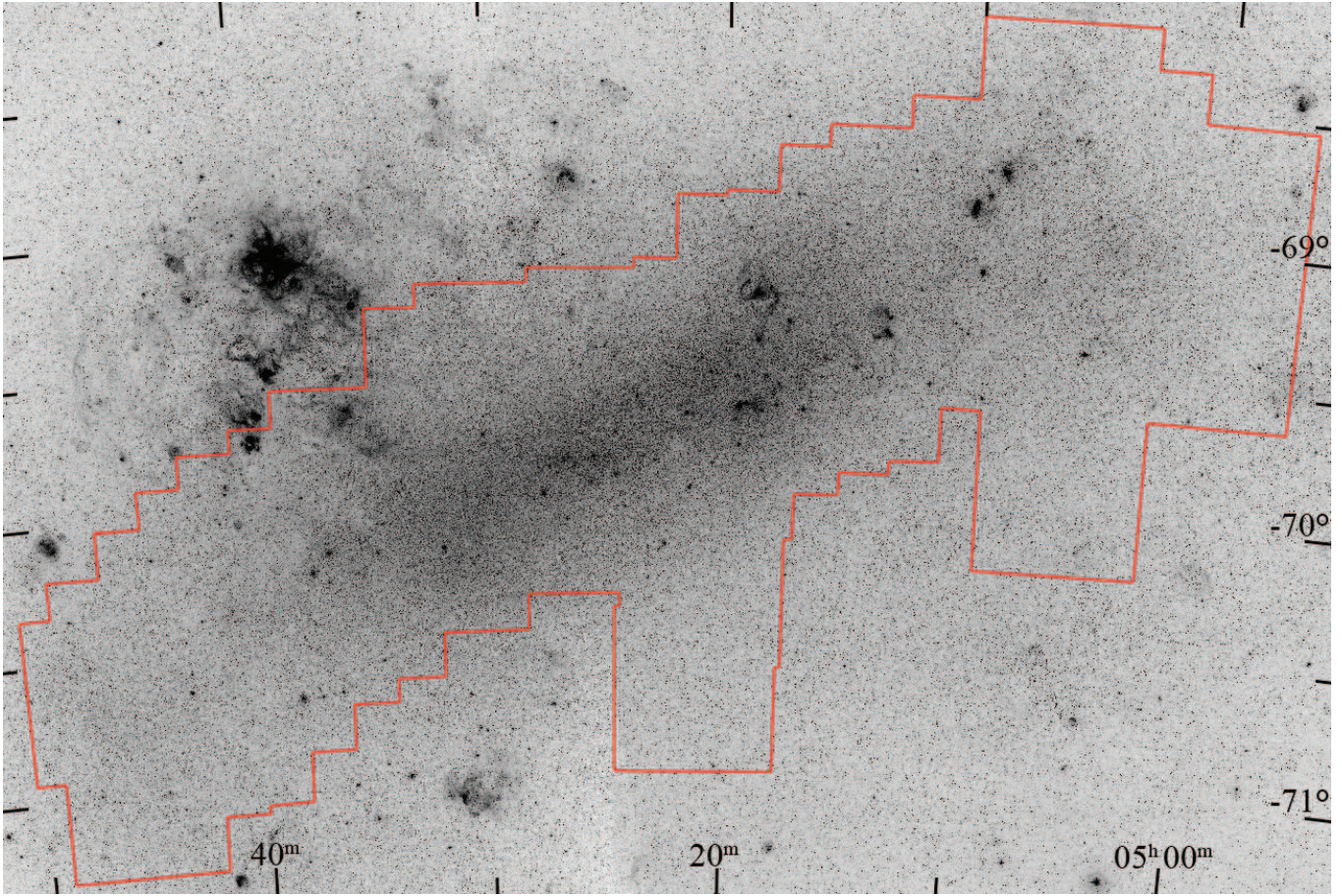}
\caption{\label{fig:dss} Digitized Sky Survey image of the Large Magellanic Cloud showing the area covered by our CPAPIR observations (red outline), which amounts to 18~sq.~deg.}
\end{center}
\vspace{-12pt}
\end{figure*}

Images were acquired using the CPAPIR camera \citep{artigau04} at the 1.5-m telescope of the Cerro Tololo Inter-American Observatory, operated by the SMARTS consortium. CPAPIR uses a $2048\times2048$ Hawaii-2 infrared array detector and delivers an effective plate scale at this telescope of $0\farcs983/$pix, or a field of view of $0\fdg559$ on a side. We requested observations centered on 49 different positions, with extensive overlap among neighboring fields to enable a robust photometric cross-calibration. Fig.~\ref{fig:dss} shows the area covered by the observations, which amounts to slightly over 18~sq.~deg. 

\begin{figure}
\begin{center}
\includegraphics[width=0.8\columnwidth,angle=270]{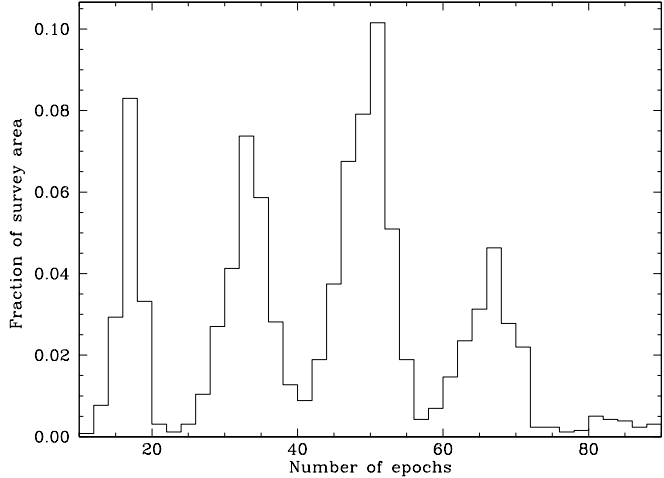}
\caption{\label{fig:epochs} Histogram of the fraction of the survey area that was imaged on a given number of epochs. The peak near 16 epochs corresponds to areas that do not overlap with neighboring fields.}
\end{center}
\end{figure}

Observations were obtained in queue mode on 32 separate nights during three distinct time periods: 2006 November (7 nights), 2007 January (6 nights) and 2007 November (19 nights). Individual fields were targeted on $7-11$ nights, often twice on each night, so that every location within our survey area was observed on $14-20$ distinct epochs. Given the significant overlap between fields, many locations within our survey area were imaged on $2-4\times$ as many epochs (see Fig.~\ref{fig:epochs}). Each unit of imaging consisted of a 6-point dither ($2\times3$ in R.A. \& decl., respectively) with commanded steps of $10\arcsec$ in each direction. A single 10s exposure was obtained at each dither point for the $J$ and $H$ sequences, while a $2\times5$s coadd was executed for the $K_s$ sequence. Calibration images (darks and dome flats in $J$ and $H$) were obtained nightly.

Images were processed using the IRAF\footnote{IRAF is distributed by the National Optical Astronomy Observatory, which is operated by the Association of Universities for Research in Astronomy (AURA) under cooperative agreement with the National Science Foundation.} packages {\tt XDIMSUM} and {\tt CCDRED}. The reduction steps consisted of bad-pixel masking (based on the median dark frame), dark current subtraction, and flat-fielding. We used dome flats for $J$ and $H$ and sky flats for $K_s$. The latter were generated by median-combining all the science images obtained on a given night, after masking all $>3\sigma$~sources present on each image. This step was performed using the {\tt xslm} routine and was repeated twice to ensure all the significant sources were masked, using the first-pass masked images for the second iteration. The reduced data set consists of 19,604 scientifically useful images, which are available upon request.

\subsection{PSF photometry and relative calibration \label{sc:psfcal}}

We performed time-series photometry of the reduced images using {\tt DAOPHOT/ALLSTAR} \citep{stetson87} and {\tt ALLFRAME}\citep{stetson94}, as well as supporting programs kindly provided by P.~Stetson. We performed the steps described below for each combination of field and filter (i.e., $49\times3$ separate reductions).

We first identified all $>5\sigma$ sources present in each image and obtained aperture photometry with a radius of 5 pixels, with a sky annulus extending from 5 to 8 pixels. We then selected up to 200 bright and isolated stars in each image to determine its point-spread function, which was modeled as a Gaussian or a Moffat profile with no spatial variation, and performed PSF photometry of all the sources detected in the first step. We used this photometry to derive accurate frame-to-frame coordinate transformations using {\tt DAOMATCH} and {\tt DAOMASTER} \citep{stetson93}. We found the typical displacement of the initial central position at each epoch to be comparable to the dither pattern (i.e, $20-30\arcsec$ or $\sim 1\%$ of the field width).

Next, we used Stetson's {\tt MONTAGE} program to select the 50 highest-quality images (using a metric based on the FWHM of the PSF and the flux level of each image) and median-combined them into a higher SNR reference frame. We repeated the steps described in the previous paragraph for the reference frame and used the resulting star list as input for the {\tt ALLFRAME} PSF photometry run. Once this was completed, we determined a position-dependent magnitude correction for each image, relative to the photometry of the frame with the highest value of the quality metric, as follows. We used $\sim 1000$ bright stars distributed throughout the field to calculate the mean magnitude offset as a function of position, carrying out the calculation every 45~pix in $x$ \& $y$ based on stars within $\pm180$~pix, and fit a thin-plate spline to the resulting values. We applied the magnitude correction by evaluating the spline fit at the position of each star. This procedure was very effective at removing low-frequency spatial variations in the photometric zeropoint, specially near the edges of the field where the optical distortion of the camera is more pronounced. As an example, the $J$-band images of field 1 typically exhibited zeropoint variations of 0.08 \& 0.22~mag (50\% and 90\% widths of the distribution, respectively) before the correction and 0.02 \& 0.06~mag after the correction. Next, we used Stetson's {\tt TRIAL} program to extract light curves and to calculate mean instrumental magnitudes and variability indices \citep[$J_{\rm Stet}$,][]{stetson96}. Fig.~\ref{fig:photint} shows our internal photometric precision as a function of magnitude in each of the three bands, for stars without any discernible intrinsic variability.

\subsection{Calibration to the 2MASS system\label{sc:cal}}

Once the preceding steps were completed, we matched and merged the final $J$, $H$ and $K_s$ star catalogs of each field using {\tt DAOMATCH} and {\tt DAOMASTER} and carried out the final astrometric and photometric calibrations using the 2MASS Point Source Catalog \citep{cutri03a} as reference. The calibrations were derived and applied separately for each of the 49 fields. Regions in common between neighboring fields were later used to test the quality of these calibration procedures.

We performed the astrometric calibration using {\tt WCSTools} \citep{mink02}, based on $600-900$~bright stars in common between 2MASS and the star catalog of a given field. Once a solution was determined and applied for a given field, we matched its full photometric catalog against 2MASS with a radial tolerance of $1\arcsec$ and found $6-14\times 10^3$~stars in common. Based on the distribution of the residuals, we determined an astrometric uncertainty of $0.15\arcsec$ (all uncertainties quoted in this work are $1\sigma$ values).

\begin{figure}
\begin{center}
\includegraphics[width=\columnwidth]{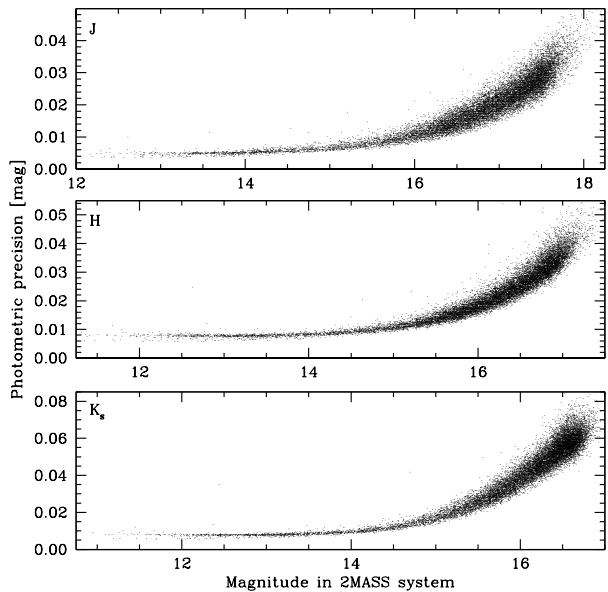}
\caption{\label{fig:photint} Internal photometric precision of our observations as a function of magnitude for the $J$ (top), $H$ (center) and $K_s$ (bottom) bands for one representative field. We selected stars with three-band photometry, Stetson variability index $J\leq0.5$ in all bands, and located within $1000$~pix in radius of the field center.}
\end{center}
\end{figure}

The absolute photometric calibration was carried out using the same type of spline-fitting procedure described in \S\ref{sc:psfcal}. While that procedure only corrected the spatial variations in the zeropoint of a given frame to the instrumental system of its reference frame, this step corrected the ensemble photometry of each field/filter combination for the spatial variations in the zeropoint of its reference frame. We used $2-8\times 10^3$~stars (depending on field and filter) for this correction, limiting the sample to objects with 2MASS magnitudes fainter than 11 (to avoid nonlinearities) and 2MASS photometric uncertainties below 0.1~mag, equivalent to $J\sim 16.1$, $H\sim 15.2$ and $K_s\sim 14.6$~mag (as shown in Fig.~\ref{fig:photint}, the CPAPIR internal photometric uncertainties were $5-15\times$ smaller). As in the previous case, the use of a spline-fitting technique was very helpful in decreasing the dispersion in the photometric solution, achieving reductions of a factor of $\sim1.5$ in $J$ and $H$ and $\sim2$ in $K_s$.

\begin{figure}
\begin{center}
\includegraphics[width=0.8\columnwidth,angle=270]{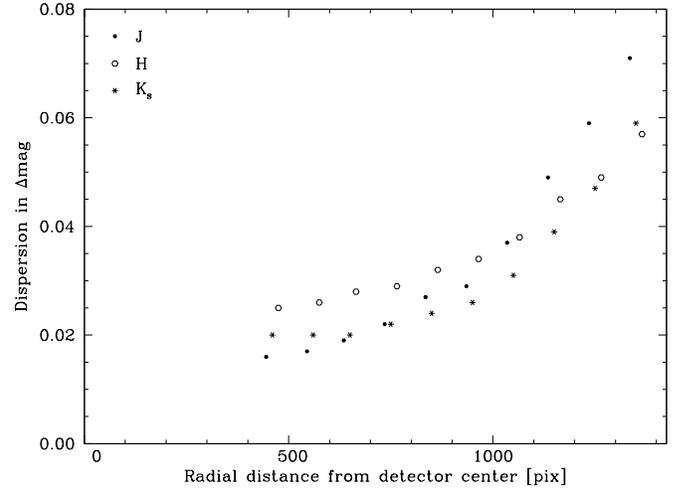}
\caption{\label{fig:photext} Test of our external photometric calibration uncertainty as a function of radial distance from detector center for the $J$ (filled circles), $H$ (open circles) and $K_s$ (stars) bands. The lack of measurements at radial distances below 400~pix is due to the fact that even the most overlapping fields were offset by at least $1/4$ of the detector width. Note the rapid degradation of photometric accuracy toward the corners of the detector.}
\end{center}
\end{figure}

We quantified the accuracy of our absolute photometric calibration by identifying objects in common between overlapping fields and calculating the uncertainty in $\Delta$mag for stars with the highest internal precision ($J\leq 15$, $H \leq 14.5$, $K_s \leq 14$) as a function of radius from the field center. The results of this comparison are plotted in Fig.~\ref{fig:photext}, with the lack of stars at radii below 400~pix due to the fact that even the most overlapping fields were offset by at least $1/4$ of the detector size. It can be seen that our total photometric calibration uncertainty near the center of the detector is only 11, 18 and 14~mmag in $JHK_s$, respectively, degrading rapidly for stars located at radii beyond the width of the detector. This position-dependent uncertainty has been fully propagated.

As part of our photometric calibration procedure, we also solved for color terms such as
\vspace{-6pt}
\begin{equation}
J\!=\!J' + \xi^J_{JK_s} (J - K_s)
\end{equation}
\noindent{where $J$ and $K_s$ are the fully-calibrated magnitudes in the 2MASS system, $J'$ is the CPAPIR magnitude partially calibrated into the 2MASS system (zeropoint-corrected but not color-term corrected), and $\xi^J_{JK_s}$ is the color term for $J$ based on the $J-K_s$ color. This was, in fact, the only statistically significant color term that we determined, with a value of $+0.018\pm0.002$.}

\subsection{Crowding corrections\label{sc:crowd}}

Even though our fields do not have a very high density of resolved point sources ($6\times 10^{-3}$ to $4\times 10^{-2}$ stars/pixel) we carried out a full suite of simulations to characterize any photometric biases due to crowding. We used the {\tt ADDSTAR} routine in {\tt DAOPHOT} to randomly place artificial stars in all master images using their corresponding PSFs, subject to the following constraints: (i) the number of artificial stars added was only 5\% of the number of actual point sources (so as to not excessively increase the stellar density of each field); (ii) the magnitudes of the artificial stars were obtained by randomly sampling the observed luminosity function; (iii) no artificial star was allowed to fall within 2.5 pixels of any other object (real or artificial) to avoid blends; (iv) the positions of the artificial stars were restricted to the innermost 80\% of the detector in $x\&y$, to avoid the PSF distortion near the edges of the field; (v) 20 realizations of each master frame were produced, to increase the statistics.

\begin{figure}
\begin{center}
\includegraphics[width=\columnwidth]{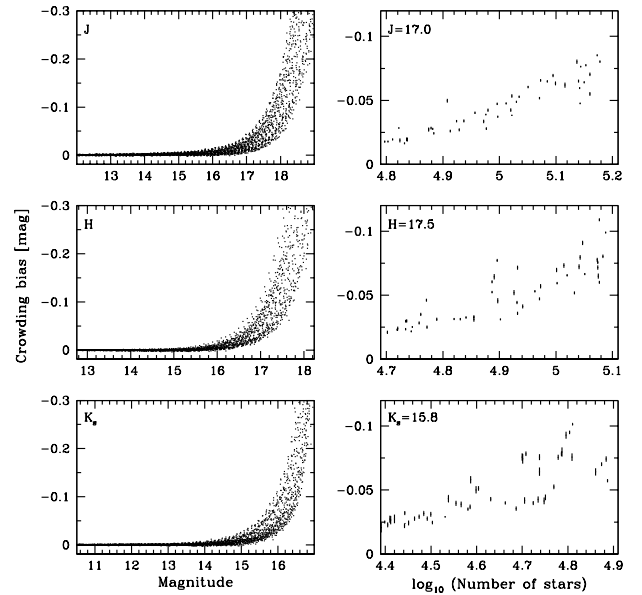}
\caption{\label{fig:crowd} Results of the artificial star simulations to characterize photometric bias due to crowding. Left: bias versus magnitude for each of the 49 fields. Right: The spread in bias at a given magnitude is well correlated with the total stellar density of the field.}
\end{center}
\end{figure}

We carried out photometry of the fake master images using {\tt ALLSTAR} in the same manner as when we analyzed the original master images, matched the input and output star lists and derived offsets as a function of magnitude shown in the left panels of Fig.~\ref{fig:crowd}. The spread in crowding bias at the faint end for a given magnitude is well correlated with the total number of stars present in each field, as seen on the right panels of that Figure. 

 We determined crowding corrections for every star in our catalog by fitting the bias-magnitude relation for its given field and filter over the magnitude range $-1 < m < 0.15$ with a fourth-degree polynomial. We removed from our catalog magnitude measurements that would have required a crowding correction exceeding 0.25~mag, as the uncertainties for larger corrections cannot be reliably determined. The crowding bias is negligible for our field standards and only amounts to a few mmag for the overwhelming majority of the Cepheids (see \S\ref{sc:lcs}). 

\vfill

\section{Results}

\subsection{Ensemble photometry\label{sc:phot}}

\begin{figure*}
\begin{center}
\includegraphics[height=\textwidth,angle=90]{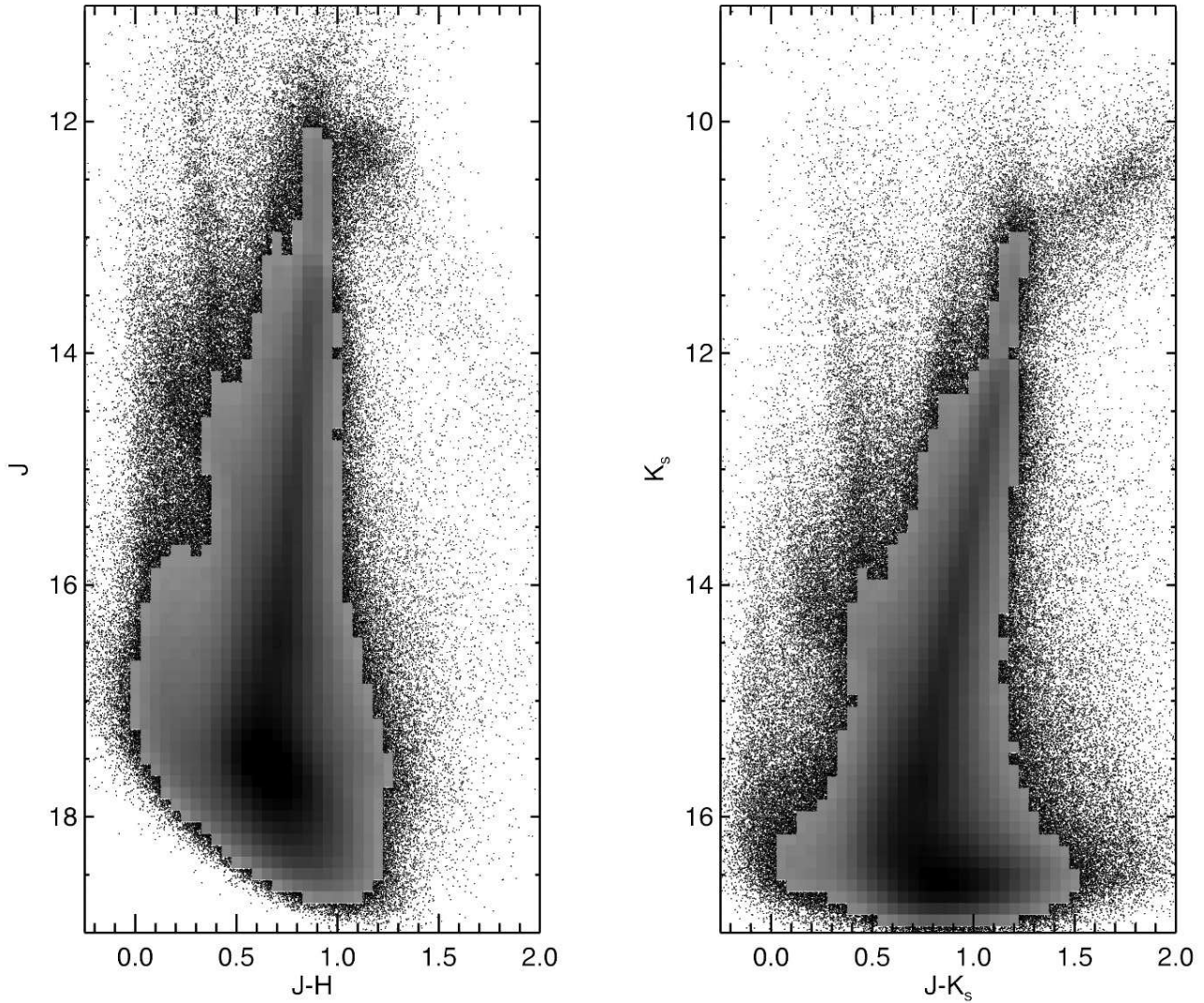}
\caption{\label{fig:cmds} Color-magnitude/Hess diagrams based on $\sim 3.6\times 10^6$ (left) and $2\times 10^6$ (right) stars with a minimum of two-band photometry. The Hess diagram is used in areas where the stellar density exceeds 200 objects per bin.}
\end{center}
\end{figure*}

\begin{figure}
\begin{center}
\includegraphics[width=\columnwidth]{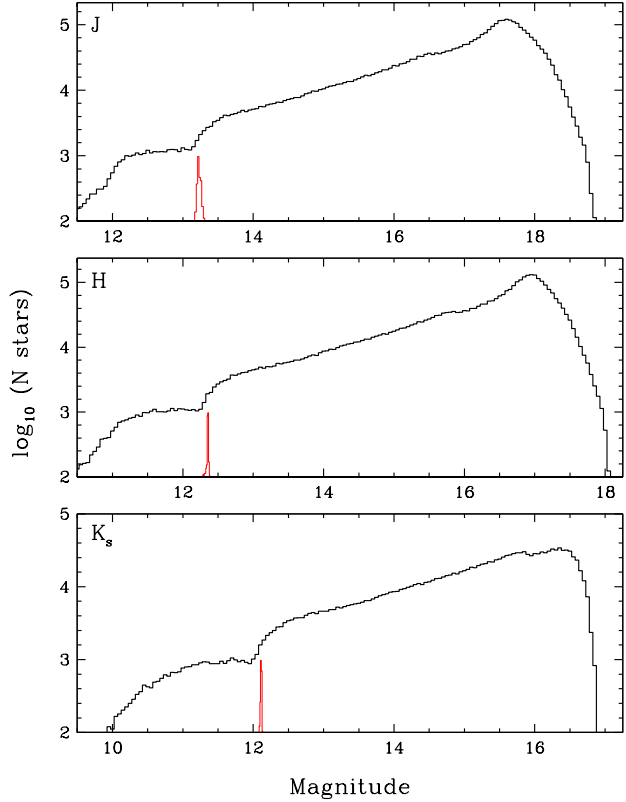}
\caption{\label{fig:lfs} Differential luminosity functions in $J$ (top), $H$ (center) and $K_s$ (bottom) for all stars in our survey area with $J-H>0.4$ (top \& center) or $J-K_s>0.5$ (bottom). The histograms plotted in red show the distribution of TRGB magnitudes obtained in 400 realizations of the edge-detection filter algorithm described in \S\ref{sc:phot}.}
\end{center}
\end{figure}

While the primary goal of this survey was to obtain well-sampled light curves of Cepheid variables, the ensemble photometry obtained as a byproduct can be used to characterize other distance indicators. The coordinates, fully-calibrated mean magnitudes and variability indices of stars with photometry in at least two bands are listed in Appendix Table~\ref{tb:phot}. The color-magnitude/Hess diagrams plotted in Fig.~\ref{fig:cmds} show that stars in the red giant branch (RGB) constitute the majority of the objects in our photometric catalog. The luminosity functions of all stars in our catalog with $J-H>0.4$~mag (for $J$ \& $H$) or $J-K_s>0.5$ (for $K_s$) are plotted in Fig.~\ref{fig:lfs}, indicating a clear detection of the Tip of the RGB (TRGB) at all wavelengths and the red clump (RC) in $J$ and $H$.

 We determined the TRGB magnitudes by selecting stars over a narrower color range ($0.75 < J-H < 1.0$ for the $J$ \& $H$ LFs, and $0.95 < J-K_s < 1.3$ for the $K_s$ LF), calculating Gaussian-smoothed luminosity functions as defined in Equation A1 of \citet{sakai96} (setting a minimum value of $\sigma_i\!=\!0.01$~mag) and applying the modified Sobel edge-detection filter as implemented in Equation 4 of \citet{mendez02} (with $\bar{\sigma}_m\!=\!0.025$~mag). The parameter choices were motivated by the very small (few mmag) internal measurement uncertainties for stars near the TRGB. We identified the local maximum of the edge-detection function in the vicinity ($\pm 0.3$~mag) of the TRGB and fit it with a spline to determine its peak value. We characterized the uncertainty in our measurements by carrying out 400 bootstrap realizations of this procedure, in which the magnitudes were perturbed by their errors. We obtained TRGB magnitudes of $J\!=\!13.23\pm0.03, H\!=\!12.35\pm0.02$ and $K_s\!=\!12.11\pm0.01$~mag. 

\begin{figure}
\begin{center}
\includegraphics[width=\columnwidth]{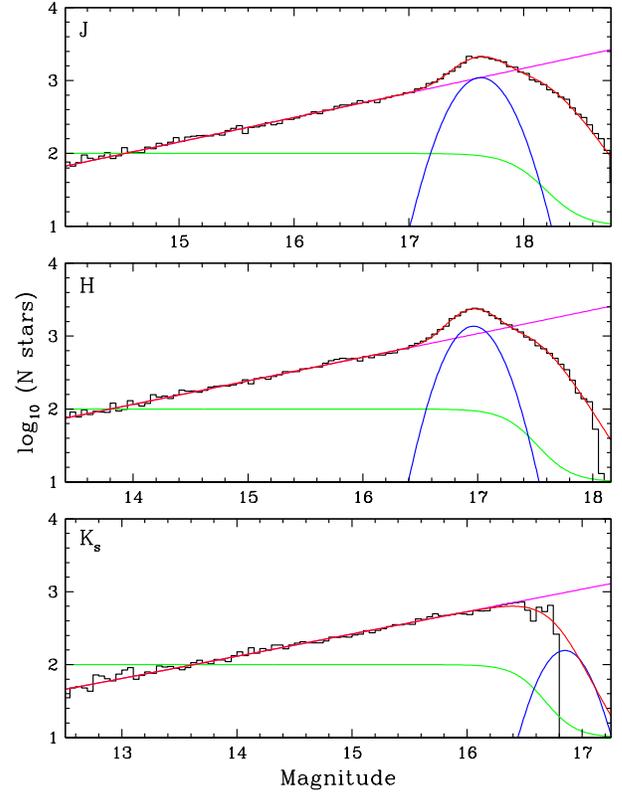}
\caption{\label{fig:rc} Differential luminosity functions for a typical field in $J$ (top), $H$ (center) and $K_s$ (bottom) for stars with $0.4 < J-H < 1.0$ (top \& center) or $0.5 < J-K_s < 1.1$ (bottom), fit using Equation~\ref{eqn:lf}. The magenta line represents the power-law component that fits the RGB LF, the blue line indicates the Gaussian component that fits the red clump, and the green line shows the photometric incompleteness function $I(m)$ (offset by $+1$ for plotting purposes). The red line denotes the combination of all components.}
\end{center}
\end{figure}

We determined the RC magnitudes by fitting the LFs of stars with $0.4<J-H<1.0$~mag using the following function:
\begin{eqnarray}\label{eqn:lf}
\Phi(m) & = & I(m)\ ( 10^{\ a+bm}\ + \\ \nonumber
        &   & c\ \exp\{-(m-m_{RC})^2/(2 \sigma_{RC}^2)\} ),
\end{eqnarray}
\noindent where $a$ and $c$ are normalization factors, $b$ is the slope of the RGB LF, $m_{RC}$ and $\sigma_{RC}$ are the mean magnitude and Gaussian width of the red clump, respectively. $I(m)$ models the photometric incompleteness as a function of magnitude,
\begin{equation}
I(m) = 1 / (1 + \exp \{(m-m_I)/\xi_I\}),
\end{equation}
\noindent where $m_I$ is the magnitude at which the incompleteness reaches 50\% and $\xi_I$ gives the scale length of the incompleteness cutoff. Fig.~\ref{fig:rc} shows a typical fit to the luminosity functions for one of our fields, and Table~\ref{tb:rc} summarizes the mean values of the various parameters. We found $\langle m_{RC} \rangle\!=\!17.60\pm0.03$ \& $16.95\pm0.04$~mag in $J$ \& $H$, respectively. We were unable to determine a reliable measurement of $m_{RC}$ at $K_s$ due to the severe incompleteness of our photometry and the significantly larger crowding corrections at the expected red clump magnitude ($K_s \sim 16.9$~mag).

\subsection{Cepheid sample and light curves\label{sc:lcs}}

Our Cepheid sample is based on the catalogs produced by the OGLE-III project \citep{soszynski08,ulaczyk13}, which provide very high-quality uniform optical photometry, mode classification and light curve parameters such as period and time of maximum light in $I$ (hereafter $T_{I,{\rm max}}$). We identified \ncpfu\ fundamental-mode (hereafter, FU) and \ncpfo\ first-overtone (hereafter, FO) variables within our survey area, covering the period range of $1.14 < P < 52.9$~d for FU and $0.27 < P < 5.91$~d for FO. The fully-calibrated light curve data for all Cepheids (including photometric zeropoints, color terms and crowding corrections) is provided in Table~\ref{tb:lcs}. The crowding corrections for the FU variables were no larger than 7~mmag in $J$ and $H$, and only greater than 10~mmag for 3\% of the objects in $K_s$. The corrections were slightly more significant for the fainter FO variables, exceeding 10~mmag for 4\%, 0.5\% and 9\% of the variables in $J$, $H$ and $K_s$, respectively.

\begin{figure}
\begin{center}
\includegraphics[width=\columnwidth]{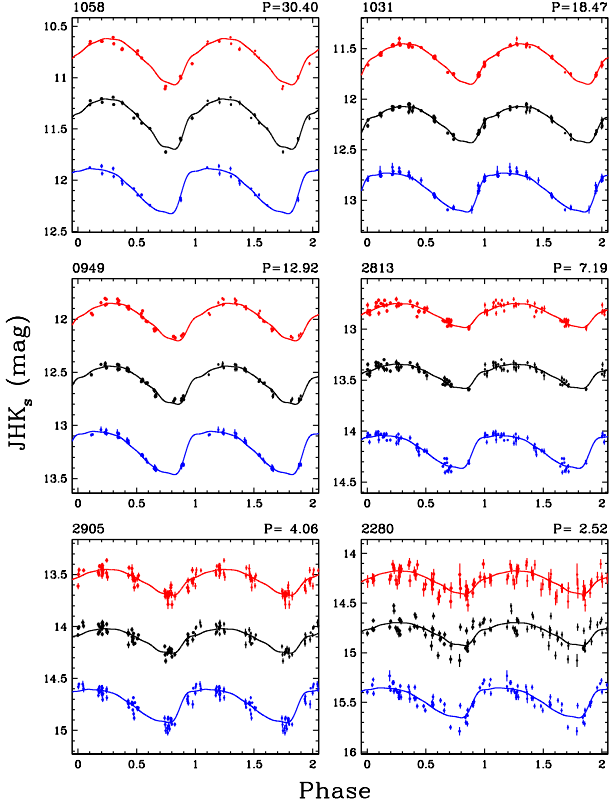}
\vspace{-18pt}
\caption{\label{fig:lcf} Representative light curves of six fundamental-mode Cepheids spanning the entire range of periods in our sample. The $J$ and $K_s$ light curves (in blue and red, respectively) have been offset for clarity by $+0.25$ and $-0.5$~mag. The solid lines represent the best-fit templates from \citet{soszynski05}.}
\end{center}
\end{figure}  

\begin{figure}
\begin{center}
\includegraphics[width=\columnwidth]{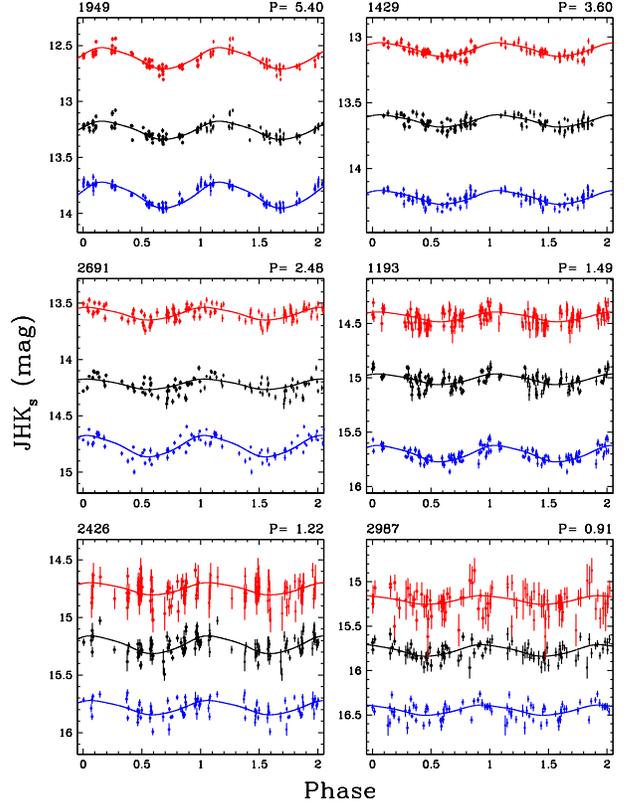}
\vspace{-18pt}
\caption{\label{fig:lco} Representative light curves of six first-overtone Cepheids spanning the entire range of periods in our sample. The $J$ and $K_s$ light curves (in blue and red, respectively) have been offset for clarity by $+0.25$ and $-0.5$~mag. The solid lines represent the best-fit sinusoidal templates.}
\end{center}
\end{figure}

We phased the Cepheid magnitudes using the periods and $T_{I,{\rm max}}$ from the aforementioned OGLE-III catalogs and fit the light curves using the templates of \citet{soszynski05} for FU variables and sinusoidal templates for the FO variables. Figs.~\ref{fig:lcf} \& \ref{fig:lco} show representative light curves and their corresponding template fits for FU and FO variables, respectively. We independently solved for the light curve amplitude in each band and we solved for a common phase offset between maximum light in $JHK_s$ and $I$ (hereafter,$\Delta\phi(I,JHK_s)$). We calculated mean magnitudes through numerical integration of the best-fit templates and estimated the magnitude uncertainties from the {\it rms} of the light curve data about the template. Table~\ref{tb:mag} lists the Cepheid properties, along with individual reddening values \citep[obtained from the extinction map of][]{haschke11}, a quality flag, and a flag to identify variables that were used in our final Leavitt Law fits (see \S\ref{sc:plr} for details).

\begin{figure}
\begin{center}
\includegraphics[width=\columnwidth]{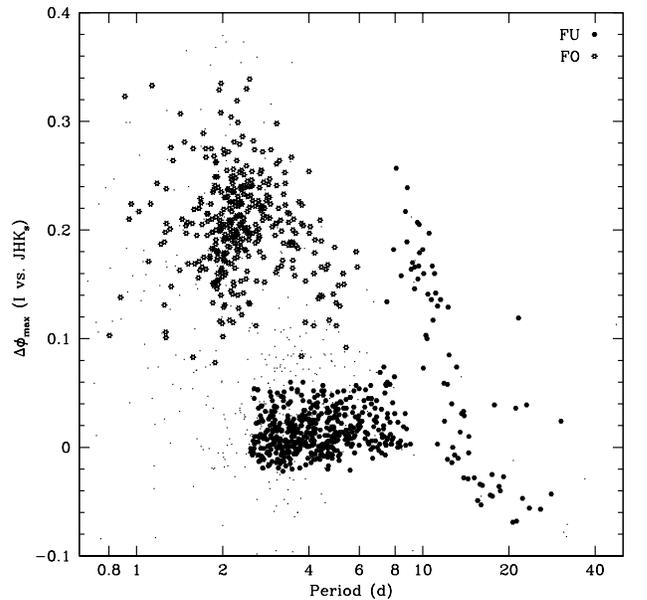}
\vspace{-18pt}
\caption{\label{fig:phd} Phase difference for maximum light in $JHK_s$ versus $I$ for the Cepheids in our sample with the highest-quality light curves (filled and open symbols represent fundamental-mode and first-overtone variables, respectively, while small dots represent variables with lower-quality light curves.}
\end{center}
\end{figure}

We classified our light curves into several quality bins, listed in Table~\ref{tb:qf} to later investigate any possible influence in our fits. Fig.~\ref{fig:phd} highlights the variation in $\Delta\phi(I,JHK_s)$ versus $\log P$ for the highest-quality variables in the FU and FO samples. We observe a mild dependence of this parameter with period for FU variables with $P<8$~d, and the gap due the Hertzsrpung progression is nicely detected. This topic is further explored in \citet{bhardwaj15}.

As an external check of our procedures for photometric calibration and determination of mean magnitudes, we compared the values obtained for 23 Cepheids in common with P04. We transformed their magnitudes using the relations between the LCO and the 2MASS systems given in \S VI.4.b of \citet{cutri03b}:
\begin{eqnarray}
K_{s,2}\!-\!K_{s,L} & = & 0.002(J\!-\!K_s)_L - 0.015, \nonumber\\
(J\!-\!K_s)_2     & = & 1.012(J\!-\!K_s)_L - 0.007, \nonumber\\
(H\!-\!K_s)_2     & = & 1.015(H\!-\!K_s)_L + 0.003, \nonumber
\end{eqnarray}

\noindent{where the ``2'' and ``L'' subscripts refer to 2MASS and LCO magnitudes, respectively. We found offsets (P04$-$this work) of $-10\pm9$, $8\pm8$ and $3\pm6$~mmag in $JHK_s$, respectively.}

\begin{figure*}
\begin{center}
\includegraphics[width=\textwidth]{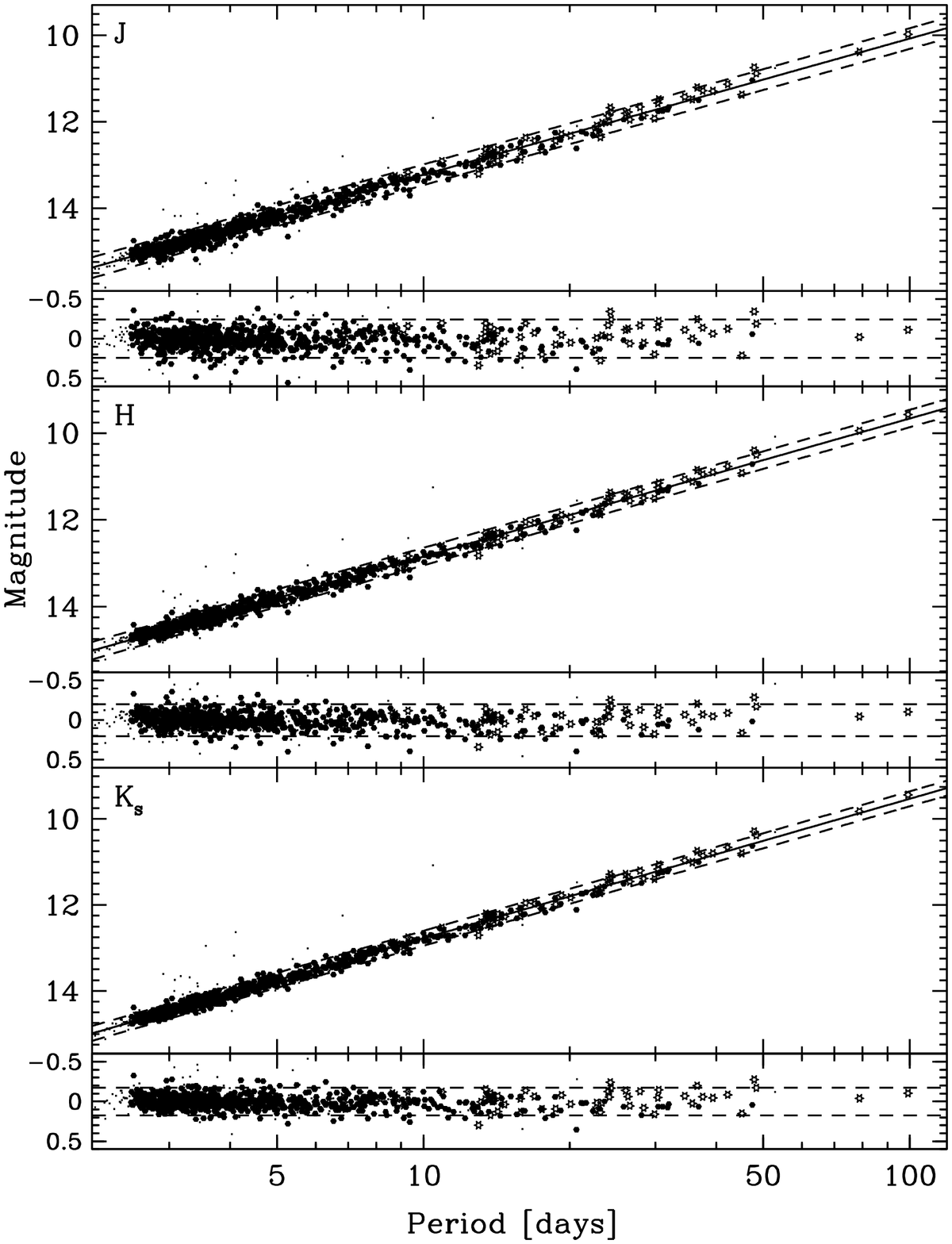}
\vspace{-9pt}
\caption{\label{fig:plf} Leavitt Law and residuals in $J$ (top), $H$ (middle) and $K_s$ (bottom) for FU Cepheids, based on \ncpfu\ objects from our sample (filled symbols) and \ncper\ variables from \citet{persson04} (open symbols). Outliers identified through an iterative rejection process based on correlated residuals are plotted using small dots. Dashed lines indicate the $\pm2\sigma$ widths of the relations.}
\end{center}
\end{figure*}  

\begin{figure*}
\begin{center}
\includegraphics[width=\textwidth]{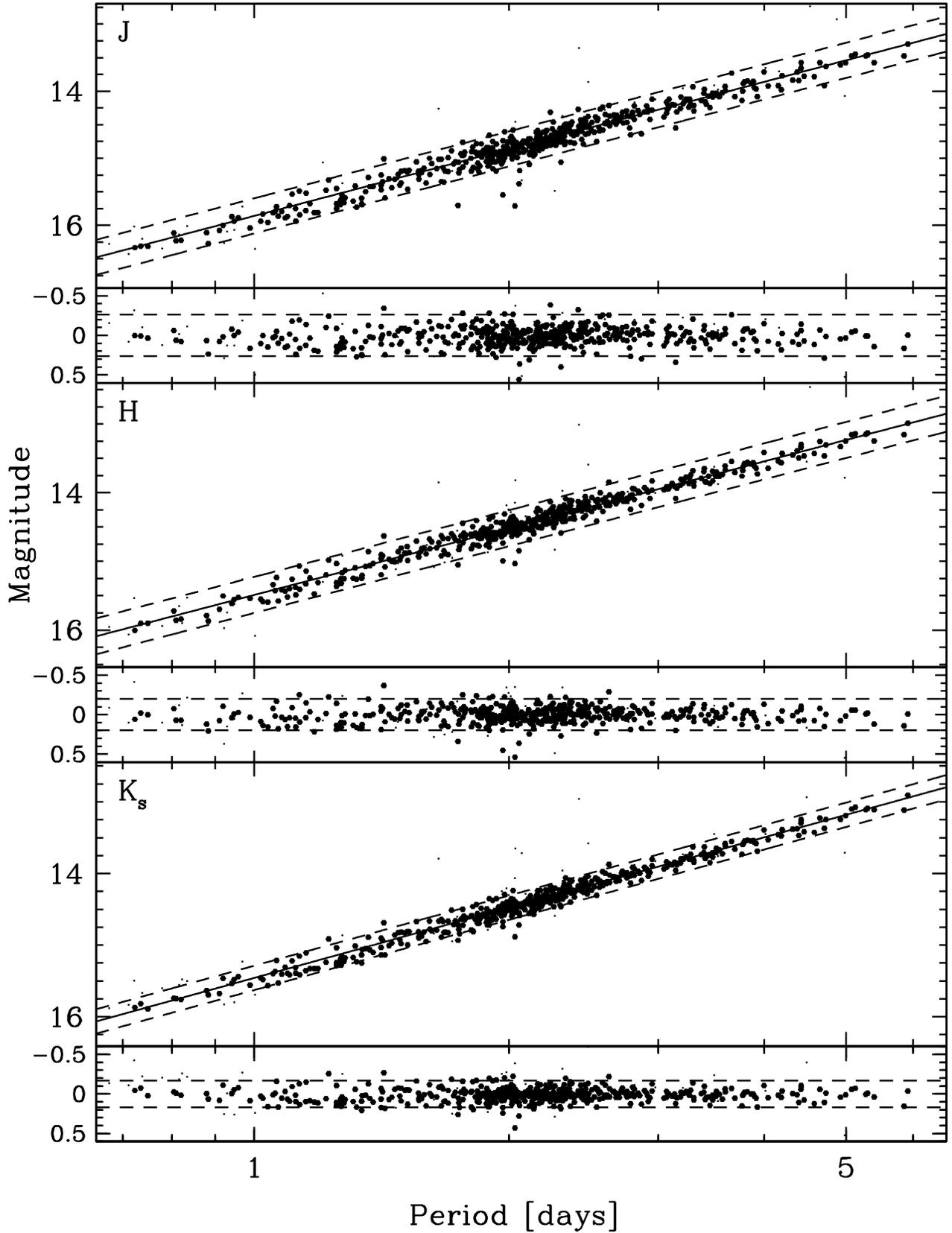}
\vspace{-9pt}
\caption{\label{fig:plo} Leavitt Law and residuals in $J$ (top), $H$ (middle) and $K_s$ (bottom) for FO Cepheids, based on \ncpfo\ objects from our sample (filled symbols). Outliers identified through an iterative rejection process based on correlated residuals are plotted using small dots. Dashed lines indicate the $\pm2\sigma$ widths of the relations.}
\end{center}
\end{figure*}

\clearpage

\subsection{Leavitt Laws\label{sc:plr}}

We corrected the mean magnitudes listed in Table~\ref{tb:mag} for the effects of interstellar dust using individual reddening values derived from the map of \citet{haschke11} and the extinction law of \citet{fitzpatrick99} applicable to the LMC. We adopted $R_V\!=\!3.1$, which for this extinction law yields ratios of total-to-selective absorption of $A_J\!=\!0.5856$, $A_H\!=\!0.3723$ \& $A_{Ks}\!=\!0.2425$ per mag of $E_{V\!I}$.

We increased the period range of the sample by adding $\ncper$ fundamental-mode variables with $P<100$~d from P04 that were not already in common with our catalog. Their magnitudes were transformed into the 2MASS system using the aforementioned relations and corrected for extinction using the same procedure as above. For those variables lying outside of the extinction map of \citet{haschke11}, we used the $E(B\!-\!V)$ values tabulated in P04. For completeness, the properties of these variables are listed in Table~\ref{tb:p04}.
 
We solved for Leavitt Laws and Period-Luminosity-Color relations of the form
\begin{eqnarray}
m & = & a + b (\log P - 1) \\
m & = & a + b (\log P - 1) + c (J\!-\!K_s)
\end{eqnarray}
\noindent following two basic assumptions: (i) the relations obey a single slope $b$ over the period range being considered; (ii) the residuals of any given Cepheid about the best-fit Leavitt Law should exhibit a strong correlation arising as a consequence of (a) uncorrected extinction or line-of-sight depth effects and/or (b) the intrinsic width in temperature of the instability strip and the resulting variation in luminosity as a function of temperature for a fixed period \citep[for a comprehensive review, see][]{madore91}.

\begin{figure}
\begin{center}
\includegraphics[width=\columnwidth]{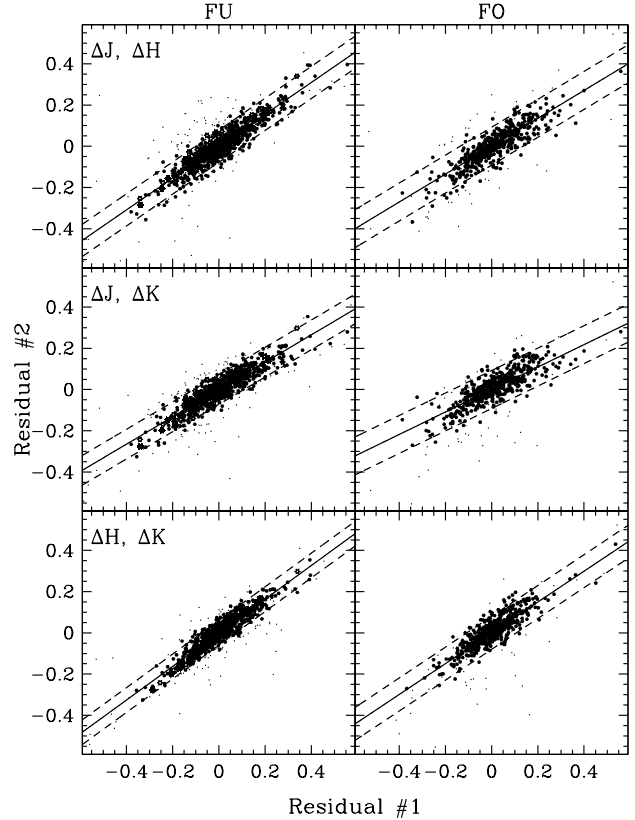}
\vspace{-18pt}
\caption{\label{fig:plr} Residuals from the Leavitt Laws plotted in Figs.~\ref{fig:plf} (left) \& \ref{fig:plo} (right). Plotting symbols follow the same convention as the aforementioned figures.}
\end{center}
\end{figure}

\begin{figure*}
\begin{center}
\includegraphics[width=\textwidth]{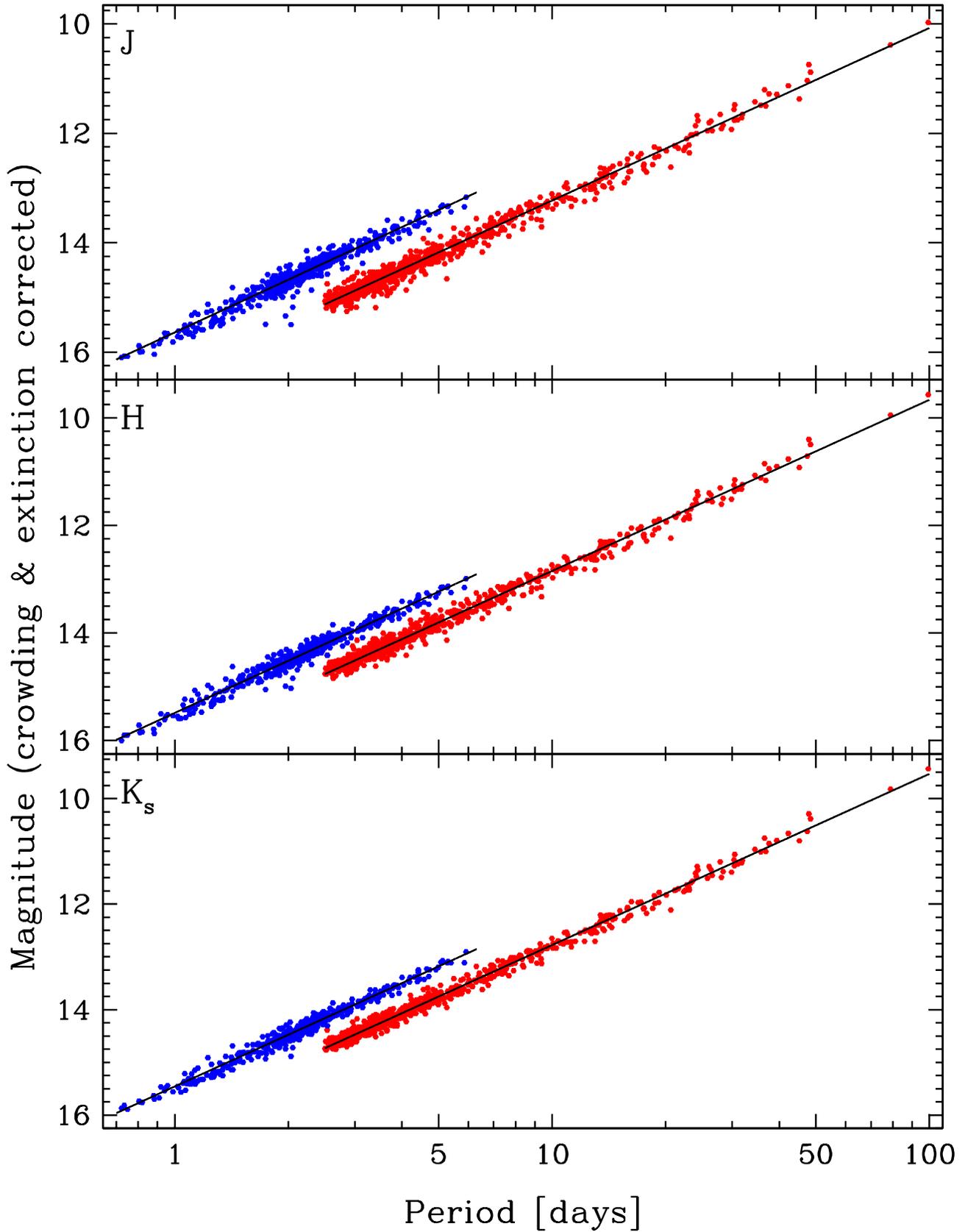}
\caption{\label{fig:plc} Final Leavitt Laws for FU and FO Cepheids, plotted using filled red and blue symbols, respectively.}
\end{center}
\end{figure*}

We started the fits using all variables except those with quality flag $G$, which are objects without three-band photometry or lying below the minimum period limits (hereafter, $P_{\rm min}$) of 0.7 \& 2.5~d for FO \& FU, respectively. We carried out the fits in an iterative manner, removing the single largest $>3\sigma$ outlier in each of the three residual relations ($\Delta J$ vs. $\Delta H$, $\Delta J$ vs. $\Delta K_s$, $\Delta H$ vs. $\Delta K_s$) until convergence. Figs.~\ref{fig:plf} \& \ref{fig:plo} show the final Leavitt Laws for FU and FO variables, respectively, while Fig.~\ref{fig:plr} shows the correlations of residuals used to identify and remove outliers. The objects included or excluded in the final fits are identified with 'Y' or 'N' in columns 17 \& 12 of Tables~\ref{tb:mag} \& \ref{tb:p04}. Fig.~\ref{fig:plc} shows the combined Leavitt Laws plotting only the Cepheids in the final samples.

\begin{figure}
\begin{center}
\includegraphics[width=\columnwidth]{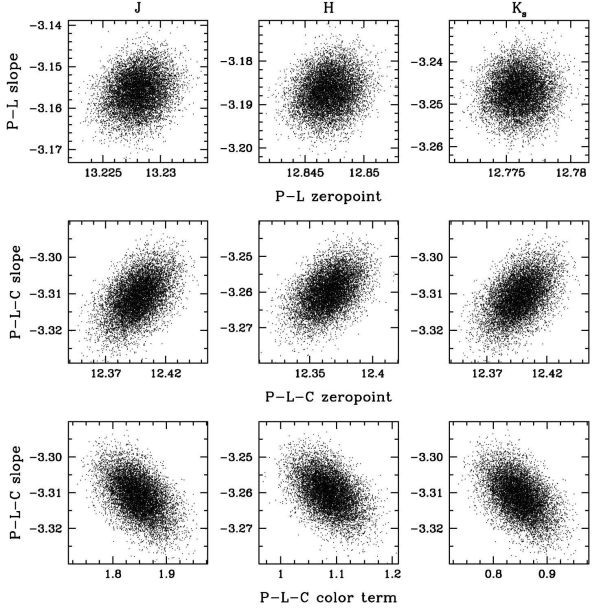}
\vspace{-18pt}
\caption{\label{fig:plpf} Result of $10^4$ random realizations of the P-L and P-L-C fitting procedures for FU Cepheids, used to estimate the statistical uncertainty and correlation of the derived parameters. The limits of each panel span $\pm4\sigma$ in the respective parameter.}
\end{center}
\end{figure}

\begin{figure}
\begin{center}
\includegraphics[width=\columnwidth]{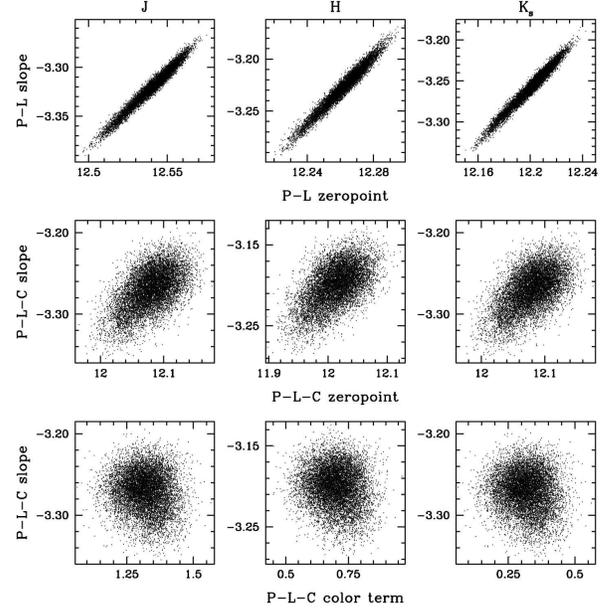}
\vspace{-18pt}
\caption{\label{fig:plpo} Result of $10^4$ random realizations of the P-L and P-L-C fitting procedures for FO Cepheids, used to estimate the statistical uncertainty and correlation of the derived parameters. The limits of each panel span $\pm4\sigma$ in the respective parameter.}
\end{center}
\vspace{-18pt}
\end{figure}

 The results of the fits are summarized in Table~\ref{tb:pls}. We estimated the statistical uncertainties in all the derived parameters by performing $10^4$ realizations of the fitting procedure in which the magnitudes were randomly altered according to their measurement errors. In order to preserve the physical correlations of the residuals, the magnitudes of a given Cepheid in all three bands were shifted using the same randomly-drawn scale factor. Figs.~\ref{fig:plpf} \& \ref{fig:plpo} show the result of this exercise for the FU \& FO samples, respectively. The much stronger correlation between parameters for the FO Leavitt Laws, relative to the FU ones, is expected given the much smaller range in period spanned by the former.  There is also a significant, but less strong, correlation between P-L-C parameters for both classes.

\ \par

Our results are in good agreement with those derived by \citet{persson04} but provide considerably stronger constraints on the slopes. Transforming the latter into the 2MASS system, we find zeropoint differences (this work-P04) of $0.061\pm0.067, 0.012\pm0.058, 0.020\pm0.054$~mag in $JHK_s$, respectively. The slopes we determined are somewhat shallower but statistically consistent given the larger uncertainties of the previous work, with differences of $-0.003\pm0.051, 0.047\pm0.042, 0.034\pm0.040$~mag/dex in $JHK_s$, respectively.

There is also very good agreement with recent theoretical calculations of the average slope over the entire period range of fundamental-mode pulsators (hereafter, $b_{\rm all}$). Following \citet{bono10}, we calculated ``LMC average'' values at $J$ and $K_s$ by combining the results for $\log(Z/X)\!=\!-2.27$ and $-1.97$ in their Table~2, and similarly at $H$ based on Table~1 from \citet{fiorentino13}. We found $b_{\rm all}\!=\!-3.15, -3.19, -3.24\pm0.05$~mag/dex in $JHK_s$, respectively, which differ by less than 0.02 from the values listed in Table~\ref{tb:pls}.

\begin{figure}
\begin{center}
\includegraphics[width=0.8\columnwidth,angle=270]{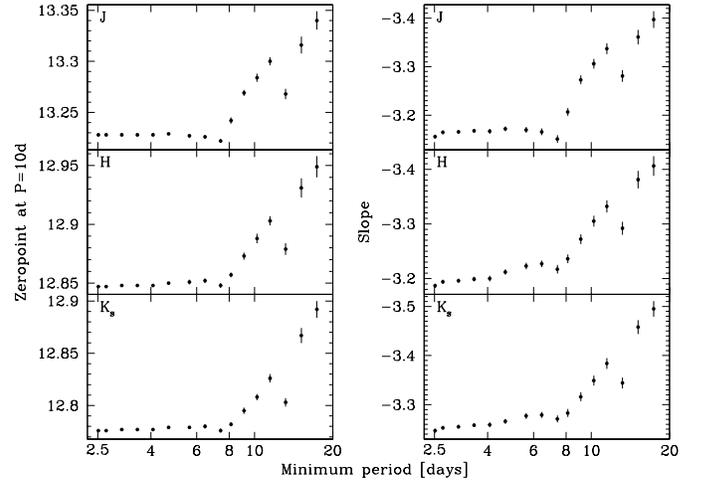}
\vspace{-6pt}
\caption{\label{fig:plcuts} Changes in Leavitt Law parameters for FU Cepheids when restricting the sample by minimum period.}
\end{center}
\vspace{-9pt}
\end{figure}

We explored the sensitivity of the derived Leavitt Law parameters to the subsample of FU Cepheids being considered, by imposing cuts based on light curve quality (Table~\ref{tb:qf}) and $P_{\rm min}$. While we found no statistically significant variation with light curve quality, there is a clear trend in the parameters with respect to the minimum period as shown in Fig.~\ref{fig:plcuts}. The theoretical expectation \citep{bono10,fiorentino13} is for the slopes to \hfill become \hfill shallower \hfill for \hfill $P_{min}>10$~d, \hfill as \hfill recently \hfill de-

\clearpage

\noindent{tected in the metal-rich Cepheids of M31 \citep{kodric15}. However we see the opposite behavior for the LMC Cepheids, with a significant increase in the slope for $P_{\rm min} \geq 8$~d. We plan further work in a companion paper to investigate this issue. The large and somewhat noisy variation in the derived values for larger $P_{\rm min}$ emphasizes the importance of obtaining large Cepheid samples to obtain robust parameters for the Leavitt Law.}

\subsection{Absolute calibration of the distance indicators\label{sc:ind}}

We derived absolute calibrations for the TRGB, red clump and the Leavitt Law using the distance to the LMC determined by \citet{pietrzynski13} using eight long-period, late-type eclipsing binary systems: $D\!=\!49.97\pm2\%$~kpc (equivalent to $\mu_0\!=\!18.493\pm0.048$~mag). We prefer this distance estimate over other contemporaneous results with slightly smaller uncertainties \citep[such as][$18.475\pm0.021$~mag]{laney12} because the method does not depend on stellar population corrections.

We corrected the TRGB magnitudes determined in \S\ref{sc:phot} for extinction using the same extinction law and total-to-selective extinction values listed above, and adopted the median reddening value in our fields from \citet{haschke11}, $\langle E_{VI} \rangle\!=\!0.08$~mag. Using the aforementioned distance modulus, we obtained $M_{\rm TRGB}\!=\!-5.31\pm0.06,-6.17\pm0.05,-6.40\pm0.05$~mag in $JHK_s$, respectively. These values are in excellent agreement with recent empirical calibrations \citep[see Fig.~5 of][]{bellazzini08} which give $M_{\rm TRGB}\!=\! -5.44,-6.30,-6.50\pm0.10$~mag for a population with the same value of ${\langle J\!-\!K_s \rangle}_{0,{\rm TRGB}}$.

 Following the same procedure for the red clump measurements, we find $M_{\rm RC}\!=\!-0.94\pm0.06$ \& $-1.57\pm0.07$~mag in $J$ \& $H$, respectively. These values can be compared with the local calibration of \citet{laney12} based on {\it Hipparcos} parallaxes, of $-0.984\pm0.014$ \& $-1.490\pm0.015$~mag. Alternatively, using the crowding- and extinction-corrected RC magnitudes and the \citet{laney12} absolute calibration we obtain an error-weighted mean LMC distance modulus of $18.49\pm0.09$~mag, which is consistent with the more precise determination by \citet{pietrzynski13}.

Lastly, using the values listed in Table~\ref{tb:pls} we find the following absolute calibration of the NIR Leavitt Laws in the LMC for fundamental-mode pulsators:
\begin{eqnarray}
J:   -5.265\pm0.049 & - & 3.156\pm0.004\ (\log P-1)  \label{eqn:plj}\\
H:   -5.646\pm0.051 & - & 3.187\pm0.004\ (\log P-1)  \label{eqn:plh}\\
K_s: -5.717\pm0.050 & - & 3.247\pm0.004\ (\log P-1), \label{eqn:plk}
\end{eqnarray}

\noindent{which includes fully-propagated uncertainties in the zeropoint due to intrinsic dispersion, photometric calibration and distance modulus.}

\section{Summary}

We have presented the details of a near-infrared ($JHK_s$) synoptic survey of the central region of the Large Magellanic Cloud, with the primary goal of providing the largest sample to date of multi-wavelength, time-resolved observations of Cepheid variables in this region of the electromagnetic spectrum. Our sample is derived from optical observations by the OGLE project \citep{soszynski08} and also benefits from an extension to longer periods by \citet{persson04}. The combined sample increases by a factor of 9 the number of available light curves with this type of data for fundamental-mode pulsators, yielding a significant increase in the accuracy with which the Leavitt Law slopes are determined. We find slopes in excellent agreement with theoretical predictions for the full period range, but we observe an unexpected steepening at long periods. We have taken advantage of the precise and accurate determination of the LMC distance using eclipsing binaries \citep{pietrzynski13} to update the absolute calibration of the Leavitt Law at these wavelengths. Furthermore, we have used our photometric database to obtain a robust absolute calibration of the Tip of the Red Giant Branch and to detect the red clump.

We plan further work based on our catalog to carry out a Fourier analysis of Cepheid light curve structure \citep{bhardwaj15}, a study of non-linearity in the Leavitt Law and P-L relations of long-period variables, among other topics.

\acknowledgements

\vspace{-12pt}

LMM acknowledges support by: NASA through Hubble Fellowship grant HST-HF-01153 from the Space Telescope Science Institute; NSF through a Goldberg Fellowship from the National Optical Astronomy Observatory and AST grant \#1211603; Texas A\&M University through a faculty start-up fund and the Mitchell-Heep-Munnerlyn Endowed Career Enhancement Professorship in Physics or Astronomy. SMK thanks SUNY-Oswego for startup funds that funded much of the initial telescope time. This research was supported by the Munich Institute for Astro- and Particle Physics (MIAPP) of the DFG cluster of excellence "Origin and Structure of the Universe". LMM \& SMK also thank the Indo-US Science and Technology Forum for supporting collaborative visits during which some of this work was completed. CCN acknowledges funding from the Ministry of Science and Technology of Taiwan under contract NSC101-2112-M-008-017-MY3. We thank Frank Ripple for assistance with the initial data reduction, and the anonymous referee for helpful comments.

This publication has made use of the following resources:

\begin{itemize}

\item observations obtained at the 1.5-m telescope of the Cerro Tololo Interamerican Observatory, operated by the SMARTS Consortium. CTIO is part of the National Optical Astronomy Observatory, which is operated by the Association of Universities for Research in Astronomy under contract with the National Science Foundation.

\item data products from the Optical Gravitational Lensing Experiment, conducted by the Astronomical Institute of the University of Warsaw at Las Campanas Observatory, operated by the Carnegie Institution for Science.

\item data products from the Two Micron All Sky Survey, which is a joint project of the University of Massachusetts and the Infrared Processing and Analysis Center at the California Institute of Technology, funded by the National Aeronautics and Space Administration and the National Science Foundation.

\item the Digitized Sky Surveys, produced at the Space Telescope Science Institute under U.S. Government grant NAG W-2166. The images of these surveys are based on photographic data obtained using the Oschin Schmidt Telescope on Palomar Mountain and the UK Schmidt Telescope. 

\item NASA's Astrophysics Data System at the Harvard-Smithsonian Center for Astrophysics.

\item The SIMBAD database and the VizieR catalogue access tool, operated at CDS, Strasbourg, France.

\item The McMaster Cepheid Photometry and Radial Velocity Data Archive, maintained by Prof.~D.~Welch at McMaster University, Canada.

\end{itemize}

{\it Facility:\ }\facility{CTIO: 1.5m}

\bibliography{macri}{}
\bibliographystyle{apj}

\clearpage

\begin{deluxetable}{lrrr}
\tablewidth{0pt}
\tablecaption{Result of fits to the RGB \& red clump\label{tb:rc}}
\tablehead{\colhead{Parameter} & \multicolumn{3}{c}{Value} \\
\colhead{} & \colhead{$J$} & \colhead{$H$} & \colhead{$K_s$}}
\startdata
RGB LF slope ($b$)         &  $0.34\pm0.01$ &  $0.33\pm0.01$ &  $0.34\pm0.01$\\
RC width ($\sigma_{RC}$     &  $0.17\pm0.01$ &  $0.17\pm0.01$ &       \nd     \\
RC mean mag.$^*$ ($m_{RC}$) & $17.60\pm0.03$ & $16.95\pm0.04$ &       \nd     \\
50\% incompl. ($m_I$)      & $18.04\pm0.09$ & $17.39\pm0.10$ & $16.51\pm0.05$\\
inc. scale length ($\xi_I$)&  $0.15\pm0.01$ &  $0.12\pm0.01$ &  $0.11\pm0.01$
\enddata
\tablecomments{$^*$: the quoted values include crowding corrections but have not been corrected for extinction (see \S\ref{sc:ind}.)}
\end{deluxetable}

\begin{deluxetable}{lcrrrr}
\tablewidth{0pt}
\tablecaption{Cepheid photometry\label{tb:lcs}}
\tablehead{\colhead{ID} & \colhead{Band} & \colhead{MJD$^*$} & \colhead{Phase$^\dagger$} & \colhead{Mag} & \colhead {$\sigma$}}
\startdata
0473 & J &  42.6048 & 0.313 & 15.132 & 0.037 \\
0473 & J &  42.7381 & 0.364 & 15.060 & 0.025 \\
0473 & J &  45.7243 & 0.497 & 15.117 & 0.017 \\
0473 & J &  45.8612 & 0.549 & 15.213 & 0.020 \\
0473 & J & 106.5305 & 0.587 & 15.151 & 0.017 
\enddata
\tablecomments{IDs are from the OGLE-III catalog \citep{soszynski08}; $^{*}$: JD-2450000; $^\dagger$: based on $P$ and $T_{I,{\rm max}}$ from the OGLE catalogs \citep{soszynski08,ulaczyk13}. Only the first five lines of the Table are presented here; the rest can be found in the online supplemental material.}
\end{deluxetable}

\begin{deluxetable}{lrlrrrrrrrrrrrrll}
\tablewidth{0pt}
\tablecaption{Cepheid properties\label{tb:mag}}
\tablehead{\colhead{ID} & \colhead{P (d)} & Cl & \multicolumn{5}{c}{Mean magnitudes}                           & \multicolumn{3}{c}{$\sigma$} & \multicolumn{3}{c}{LC amplitudes} & \colhead{$E_{V\!I}$} & \colhead{QF} & \colhead{UF} \\
                        &                 &    &    $V$     &    $I$     &    $J$     &    $H$     &   $K_s$   &   $J$   &   $H$   &   $K_s$  &    $J$    &    $H$    &   $K_s$   &                      &              & }
\startdata
0473 &  2.634 & FU & 16.324 & 15.588 & 15.109 & 14.845 & 14.624 &  80 & 116 & 116 & 238 & 288 & 283 &  90 & F & N\\
0474 &  0.816 & FO & 17.338 & 16.535 & 16.195 & 15.687 & 15.579 & 131 & 152 & 171 & 192 &  70 &  60 &  90 & F & N\\
0477 &  1.959 & FO & 16.142 & 15.456 & 15.142 & 14.678 & 14.607 &  50 &  54 &  64 & 274 &  50 &   9 & 100 & D & Y\\
0478 &  2.764 & FU & 16.155 & 15.464 & 14.979 & 14.654 & 14.609 &  56 &  73 &  78 & 344 &  45 &  91 & 110 & D & Y\\
0480 &  4.035 & FU & 16.868 & 15.755 & 15.119 & 14.503 & 14.497 &  48 &  52 &  80 & 555 & 489 & 624 & 130 & D & N
\enddata
\tablecomments{IDs, periods and VI magnitudes are from the OGLE catalogs \citep{soszynski08,ulaczyk13}. Magnitudes are corrected for crowding but not for extinction. $E_{V\!I}$ values are taken from \citet{haschke11}. Magnitude uncertainties, light curve amplitudes and reddenings are expressed in mmag. Cl: class (FUndamental or First Overtone); QF: quality flag (see Table~\ref{tb:qf}); UF: flag to indicate if the variable was used in the final P-L fits. Only the first five lines of the Table are presented here; the rest can be found in the online supplemental material.}
\end{deluxetable}

\begin{deluxetable}{lllr}
\tablewidth{0pt}
\tablecaption{Quality flags for Cepheid lightcurves\label{tb:qf}}
\tablehead{\colhead{Flag} & \colhead{Description} & \colhead{Range} & \colhead{$N$}}
\startdata
\multicolumn{4}{c}{Fundamental mode}\\
  &  & \\
G & Below minimum period or lacks 3-band data & $P<2.5$                                                     & 60 \\
  &  &  &     \\
  &  &  &     \\
F & Exceeds maximum lightcurve rms            & $\sigma(J) > 0.07$, $\sigma(H) > 0.08$, $\sigma(K_s) > 0.09$ & 71 \\
  &  &  &     \\
E & Outlier in color-color relation           & $J-K_s\!=\!0.165+0.749(J-H)$, $\Delta>0.13$                      & 18 \\
  &  &  &     \\
  &                                           & $H/J  \!=\!0.82+0.20\log P$, $\Delta>0.40$                      &    \\
D & Outlier in NIR amplitude ratios           & $K_s/J\!=\!0.80+0.26\log P$, $\Delta>0.36$                       & 77 \\
  &                                           & $H/K_s\!=\!1.03-0.08\log P$, $\Delta>0.52$                       &    \\
  &  &  &     \\
C & Phase difference of maximum light         & $\Delta\phi(I,JHK_s) \notin (-0.025,0.06)$ for $\log P<0.85$ & 44 \\
  &  &  &     \\
  &                                           & $J/I\!=\!0.67+0.17\log P+0.14(\log P)^2$, $\Delta>0.23$          &     \\
B & Outlier in NIR-to-$I$ amplitude ratios    & $H/I\!=\!0.54+0.29\log P+0.25(\log P)^2$, $\Delta>0.28$          & 32  \\
  &                                           & $K_s/I\!=\!0.52+0.34\log P+0.32(\log P)^2$, $\Delta>0.25$        &     \\
  &  &  &     \\
A & Passed all selection criteria             &                                                             & 564 \\
\tableline
\multicolumn{4}{c}{First overtone} \\
  &  &  &     \\
G & Below minimum period or lacks 3-band data & $P<0.7$                                                     &  30 \\
  &  &  &     \\
F & Exceeds maximum lightcurve rms            & $\sigma(J) > 0.09$, $\sigma(H) > 0.13$, $\sigma(K_s) > 0.2$ &  20 \\
  &  &  &     \\
E & Outlier in color-color relation           & $J-K_s\!=\!0.085+0.915(J-H)$, $\Delta>0.14$                     &  13 \\
  &  &  &     \\
  &                                           & $H/J\!=\!0.62$, $\Delta>0.26$                                  &     \\
D & Outlier in NIR amplitude ratios           & $K_s/J\!=\!0.61$, $\Delta>0.24$                                 &  84 \\
  &                                           & $H/K_s\!=\!1.07$, $\Delta>0.41$                                 &     \\
  &  &  &     \\
C & Phase difference of maximum light         & $\Delta\phi(I,JHK_s) \notin (0.05,0.35)$                    &  14 \\
  &  &  &     \\
  &                                           & $J/I\!=\!0.64$, $\Delta>0.13$                                  &     \\
B & Outlier in NIR-to-$I$ amplitude ratios    & $H/I\!=\!0.44$, $\Delta>0.15$                                  &  25 \\
  &                                           & $K_s/I\!=\!0.42$, $\Delta>0.14$                                 &     \\
  &  &  &     \\
A & Passed all selection criteria             &                                                            & 365
\enddata
\end{deluxetable}

\begin{deluxetable}{lrrrrrrrrrrcl}
\tablewidth{0pt}
\tablecaption{Additional Cepheids from \citet{persson04}\label{tb:p04}}
\tablehead{\colhead{ID} & \colhead{P (d)} & \multicolumn{5}{c}{Mean magnitudes}                           & \multicolumn{3}{c}{$\sigma$} & \colhead{$E_{V\!I}$} & \colhead{UF} & \colhead{Src}\\
                        &                 &    $V$     &    $I$     &    $J$     &    $H$     &   $K_s$   &   $J$   &   $H$   &   $K_s$  &                     &              &              }
\startdata
HV5541   &  2.682 & 15.970 & 15.350 & 14.928 & 14.658 & 14.593 &  50 &  35 &  61 &  42 & Y & S02\\
HV12225  &  3.007 & 16.160 & 15.420 & 14.950 & 14.629 & 14.537 &  16 &  16 &  27 &  42 & Y & S02\\
HV12765  &  3.429 & 15.290 & 14.670 & 14.174 & 13.904 & 13.840 &  14 &  12 &  12 &  73 & N & S02\\
HV12747  &  3.599 & 15.770 & 15.130 & 14.641 & 14.347 & 14.279 &  22 &  18 &  16 &  42 & Y & S02\\
HV12226  &  3.706 & 15.874 & 15.161 & 14.517 & 14.169 & 14.182 &  50 &  51 & 183 &  51 & Y & S14
\enddata
\tablecomments{IDs, periods and VI magnitudes are from the OGLE catalogs \citep[][abbreviated as S08, U13 \& S14, respectively]{soszynski08,ulaczyk13,soszynski14} when available or otherwise from the literature \citep[][abbreviated as M79, F85, B99, T99, S02 \& N06, respectively]{martin79,freedman85,barnes99,tanvir99,sebo02,ngeow06}. Tabulated magnitudes are not corrected for extinction. $JHK_s$ magnitudes have been transformed into the 2MASS system. $E_{V\!I}$ values are taken from \citet{haschke11}, when available. Magnitude uncertainties and reddenings are expressed in mmag. UF: flag to indicate if the variable was used in the final P-L fits. Only the first five lines of the Table are presented here; the rest can be found in the online supplemental material.}
\end{deluxetable}

\begin{deluxetable}{lllll}
\tablewidth{0pt}
\tablecaption{Leavitt Laws \& Period-Luminosity-Color relations\label{tb:pls}}
\tablehead{\colhead{Band(s)} & \colhead{Zeropoint} & \colhead{Slope} & \colhead{Color term} & \colhead{\it {r.m.s.}}}
\startdata
\multicolumn{5}{l}{Fundamental mode: $P_{\rm min}\!=\!2.5$~d, $N_{\rm start}\!=\!872$, $N_{\rm final}\!=\!775$}  \\
$J$                & $13.228\pm0.002$ & $-3.156\pm0.004$ &       \nd       & 0.120 \\
$H$                & $12.847\pm0.002$ & $-3.187\pm0.004$ &       \nd       & 0.101 \\
$K_s$              & $12.776\pm0.001$ & $-3.247\pm0.004$ &       \nd       & 0.087 \\
$J,   (J\!-\!K_s)$ & $12.397\pm0.015$ & $-3.311\pm0.005$ & $1.847\pm0.033$ & 0.080 \\
$H,   (J\!-\!K_s)$ & $12.365\pm0.014$ & $-3.260\pm0.005$ & $1.086\pm0.032$ & 0.084 \\
$K_s, (J\!-\!K_s)$ & $12.397\pm0.015$ & $-3.311\pm0.005$ & $0.848\pm0.033$ & 0.080 \\
\tableline
\multicolumn{5}{l}{First overtone: $P_{\rm min}\!=\!0.7$~d, $N_{\rm start}\!=\!521$, $N_{\rm final}\!=\!474$} \\
$J$                & $12.541\pm0.012$ & $-3.319\pm0.020$ &       \nd       & 0.131 \\
$H$                & $12.262\pm0.012$ & $-3.227\pm0.020$ &       \nd       & 0.100 \\
$K_s$              & $12.201\pm0.014$ & $-3.257\pm0.023$ &       \nd       & 0.085 \\
$J,   (J\!-\!K_s)$ & $12.079\pm0.030$ & $-3.270\pm0.024$ & $1.318\pm0.065$ & 0.080 \\
$H,   (J\!-\!K_s)$ & $12.013\pm0.032$ & $-3.200\pm0.024$ & $0.698\pm0.070$ & 0.083 \\
$K_s, (J\!-\!K_s)$ & $12.079\pm0.030$ & $-3.270\pm0.024$ & $0.316\pm0.065$ & 0.080
\enddata
\tablecomments{Quoted uncertainties in zeropoints and color terms do not include external photometric uncertainties of 11, 18 \& 14 mmag in $JHK_s$, respectively (\S\ref{sc:cal}), which should be added in quadrature to the above values.}
\end{deluxetable}

\renewcommand{\thetable}{A\arabic{table}}
\setcounter{table}{0}

\begin{deluxetable}{llrrrrrrrrrr}
\tablewidth{0pt}
\tablecaption{Ensemble photometry\label{tb:phot}}
\tablehead{\multicolumn{2}{c}{Coordinates (J2000.)} & \multicolumn{3}{c}{Magnitudes} & \multicolumn{3}{c}{$\sigma(Mag)$} & \multicolumn{3}{c}{$J_{\rm Stet}$} & \colhead{Field}\\
\colhead{R.A.} & \colhead {Dec} & \colhead{$J$} & \colhead{$H$} & \colhead{$K_s$} & \colhead{$J$} & \colhead{$H$} & \colhead{$K_s$} & \colhead{$J$} & \colhead{$H$} & \colhead{$K_s$} & \colhead{\#}}
\startdata
74.28737 & -69.54611 &   \nd  & 13.661 & 13.404 &  \nd  & 0.005 & 0.007 &   \nd  &  0.356 &  0.309 &  2 \\
74.28782 & -69.57628 & 14.930 & 14.114 & 14.028 & 0.009 & 0.008 & 0.013 &  0.372 &  0.394 &  0.384 &  2 \\
74.28901 & -69.53095 & 15.686 & 14.859 & 14.813 & 0.012 & 0.012 & 0.019 &  0.123 &  0.627 &  0.278 &  2 \\
74.28933 & -69.57615 & 16.532 &   \nd  & 15.511 & 0.045 &  \nd  & 0.050 &  0.875 &   \nd  &  0.456 &  2 \\
74.28983 & -69.49224 & 16.581 & 16.440 &   \nd  & 0.018 & 0.028 &  \nd  &  0.305 &  0.194 &   \nd  &  2
\enddata
\tablecomments{Magnitudes were corrected for crowding using the procedure detailed in \S2.4, but have not been corrected for extinction. Objects with $J_{\rm Stet} > 0.75$ are likely to be variable; mean magnitudes should be considered approximate. Only the first five lines of the Table are presented here; the rest can be found in the online supplemental material.}
\end{deluxetable}
\end{document}